\documentclass[12pt,a4paper]{article}          
\usepackage{emlines,bezier,rrussian,disser,helvet,amssymb,color,graphics} 
\usepackage[cp1251]{inputenc}  
\usepackage[T2A]{fontenc}      
\usepackage[russian]{babel}    

\begin{document}

\newcommand{\beq}{\begin{equation}}
\newcommand{\eeq}{\end{equation}}

{\Large \bf Dissipative heat engine is thermodynamically inconsistent}

\vspace{0.3cm}
\noindent
A.~M.~Makarieva, V.~G.~Gorshkov

\noindent
{\it Theoretical Physics Division, Petersburg Nuclear Physics Institute, Gatchina, St. Petersburg, Russia, elba@peterlink.ru}

\vspace{0.3cm}
{\bf Abstract.}
A heat engine operating on the basis of the Carnot cycle is considered, where the mechanical work performed is dissipated
within the engine at the temperature of the warmer isotherm and the resulting heat is added to the engine together
with an external heat input. The resulting work performed by the engine per cycle is increased at the expense
of dissipated work produced in the previous cycle. It is shown that such a dissipative heat engine is thermodynamically inconsistent
violating the first and second laws of thermodynamics. The existing physical models employing the dissipative heat engine concept, in particular, the
heat engine model of hurricane development, are physically invalid.

\vspace{0.3cm}

\section{Introduction}

The Carnot cycle does not involve irreversible processes or dissipative losses. In a Carnot
heat engine the working body (a fluid capable of expansion, usually gas) receives
heat $Q_s$ from a hot body (the heater) and performs mechanical work $A_s = Q_s$ at temperature $T_s$.
It gives heat $Q_0$ away to a cold body (the cooler) while work $A_0 = Q_0 \le Q_s$ is performed on the gas at temperature $T_0 \le T_s$.
The resulting work $A > 0$ is determined by the energy conservation law (the first law of thermodynamics) as $A = Q_s - Q_0$.
This work is performed by the working body on its environment.
Since all the processes in the Carnot cycle are reversible, entropy of the working body is conserved.
Entropy of the environment on which the work is performed does not change either. The amount of entropy $S_s = Q_s/T_s$ received from
the heater is equal to the amount of entropy $S_0 = Q_0/T_0$ given away to the cooler.
The equality $Q_s/T_s = Q_0/T_0$ that stems from the second law of thermodynamics combines
with the energy conservation law to determine efficiency $\varepsilon \equiv A/Q_s$ of the Carnot cycle as $\varepsilon = (T_s - T_0)/T_s < 1$.
For the Carnot heat engine $A \le Q_s$.

The dissipative heat engine concept advanced in [1] and discussed in [2-5] is used in modern
meteorological literature to account for hurricane intensity [6,7]. In the dissipative heat engine work $A_d$ (low index
$d$ stands for the dissipative heat engine) produced in the
ideal Carnot cycle undergoes dissipation at temperature $T_s$ of the heater. The resulting heat is added to the working body
together with the external heat $Q_{sd}$ that comes from the heater. In the stationary case the relationship between work $A_d$ and external
heat $Q_{sd}$ is then written as $A_d = \varepsilon (Q_{sd} + A_d)$. Efficiency $\varepsilon_d = A_d/Q_{sd}$ of the dissipative heat engine becomes
$\varepsilon_d = (T_s - T_0)/T_0$.  Thus, for a given $Q_{sd}$ the efficiency of the dissipative heat engine grows infinitely
with decreasing $T_0$, and work $A_d$ can become much larger than $Q_{sd}$: $A_d \gg Q_{sd}$ at $T_0 \ll T_s - T_0$ and $\varepsilon_d \gg 1$;
$A_d \to \infty$ at $T_0 \to 0$.
Demanding energy to be conserved gives $Q_{sd} = Q_0$, i.e. the amount of heat received by the dissipative heat engine from the heater
coincides with the amount of heat disposed to the cooler. It is assumed [1-7] that when work $A_d$ dissipates within the working body
of the dissipative heat engine in contact with the heater, i.e. at $T = T_s$, entropy increases by $S_{sd} = (Q_{sd} + A_d)/T_s$.
The decrease of entropy due to contact with the cooler remains $S_0 = Q_0/T_0$ as in the Carnot heat engine. Taking into account
that $Q_s = Q_0$ and $A_d = \varepsilon_d Q_{sd}$, the mathematical equality  $S_{sd} = S_0$ holds. From this it is concluded that
in the dissipative heat engine the entropy of the working body remains constant [5] and that the dissipative heat engine
conforms to both first and second laws of thermodynamics and can exist. The main feature of the dissipative heat engine is
the possibility to significantly increase work $A_d$ produced by the engine per cycle, to $A_d > Q_{sd}$ and ultimately to infinity
compared to the ideal Carnot cycle, where $A$ is always less than $Q_s$.

In this paper we show that the concept of the dissipative heat engine is based an a physical misinterpretation of the nature of
the Carnot heat engine. When the essential physical features of the heat engine are taken into account, the concept of the dissipative heat engine is shown to
be in conflict with the laws of thermodynamics.

\section{Physics of the Carnot heat engine}

The Carnot cycle consists of two isotherms at temperatures $T = T_s$ and $T = T_0$ of the heater and the cooler, respectively,
and of two adiabates connecting the isotherms. The working body in thermodynamic equilibrium with the heater at $T = T_s$ cannot receive
heat from the latter. To receive heat, the working body must expand first, so that its temperature becomes a little lower
than that of the heater, only then the heat flux from the heater to the working body becomes possible. Thus, the Carnot heat engine
operating on the basis of the Carnot cycle must be furnished with an auxilliary dynamic device that performs mechanical expansion and
contraction of the working body. A Carnot heat engine where the role of such a device is played by an ideal
elastic spring is shown in Fig.~1. The working body (gas) is contained in the cylinder capped
on one side by a sliding piston that is connected to the spring. The piston travels within the cylinder without friction.
Two stops are provided to limit the piston's movement and to define the minimum and maximum volumes occupied by the working body, Fig.~1.
In the ultimate states of maximum compression and extension of the spring its potential energy is maximum.
In the intermediate state where the spring is relaxed, its potential energy is zero, but the kinetic energy of the working body and the piston is not zero.

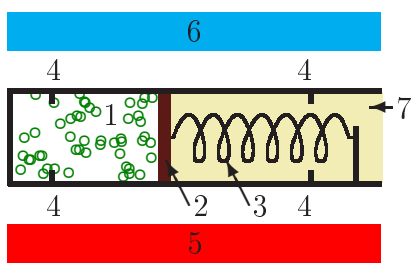
\begin{figure}
\caption{A Carnot heat engine. The working body (gas) (green balls) is contained within a closed cylinder ({\it 1})
with a sliding piston ({\it 2}) as the cap. Spring ({\it 3}) makes the piston move. Elastic stops ({\it 4}) limit the
minimum and maximum volume of the gas in the cycle. During the cycle, the working body is sequentially brought in contact
with the heater ({\it 5}) at $T = T_s$ and the cooler ({\it 6}) at $T = T_0$. The environment where the engine works ({\it 7}) is assumed
to be infinite, so its pressure $p$ is constant during the cycle.}

\newcommand{\mpt}{\multiput}

\unitlength=0.10mm
\special{em:linewidth 0.7pt}
\linethickness{0.7pt}

\begin{center}{
\begin{picture}(600.00,400.00)(0,0)

\put(0,120){

\put(50,0){

\put(88,43){\makebox(0,0)[cc]{\textcolor{protozoa}{\scriptsize \boldmath $\circ$}}}
\put(14,36){\makebox(0,0)[cc]{\textcolor{protozoa}{\scriptsize \boldmath $\circ$}}}
\put(171,55){\makebox(0,0)[cc]{\textcolor{protozoa}{\scriptsize \boldmath $\circ$}}}
\put(95,108){\makebox(0,0)[cc]{\textcolor{protozoa}{\scriptsize \boldmath $\circ$}}}
\put(174,61){\makebox(0,0)[cc]{\textcolor{protozoa}{\scriptsize \boldmath $\circ$}}}
\put(117,55){\makebox(0,0)[cc]{\textcolor{protozoa}{\scriptsize \boldmath $\circ$}}}
\put(22,31){\makebox(0,0)[cc]{\textcolor{protozoa}{\scriptsize \boldmath $\circ$}}}
\put(142,39){\makebox(0,0)[cc]{\textcolor{protozoa}{\scriptsize \boldmath $\circ$}}}
\put(102,99){\makebox(0,0)[cc]{\textcolor{protozoa}{\scriptsize \boldmath $\circ$}}}
\put(143,58){\makebox(0,0)[cc]{\textcolor{protozoa}{\scriptsize \boldmath $\circ$}}}
\put(154,12){\makebox(0,0)[cc]{\textcolor{protozoa}{\scriptsize \boldmath $\circ$}}}
\put(182,53){\makebox(0,0)[cc]{\textcolor{protozoa}{\scriptsize \boldmath $\circ$}}}
\put(11,17){\makebox(0,0)[cc]{\textcolor{protozoa}{\scriptsize \boldmath $\circ$}}}
\put(46,19){\makebox(0,0)[cc]{\textcolor{protozoa}{\scriptsize \boldmath $\circ$}}}
\put(153,84){\makebox(0,0)[cc]{\textcolor{protozoa}{\scriptsize \boldmath $\circ$}}}
\put(168,10){\makebox(0,0)[cc]{\textcolor{protozoa}{\scriptsize \boldmath $\circ$}}}
\put(41,12){\makebox(0,0)[cc]{\textcolor{protozoa}{\scriptsize \boldmath $\circ$}}}
\put(150,21){\makebox(0,0)[cc]{\textcolor{protozoa}{\scriptsize \boldmath $\circ$}}}
\put(155,106){\makebox(0,0)[cc]{\textcolor{protozoa}{\scriptsize \boldmath $\circ$}}}
\put(55,106){\makebox(0,0)[cc]{\textcolor{protozoa}{\scriptsize \boldmath $\circ$}}}
\put(144,54){\makebox(0,0)[cc]{\textcolor{protozoa}{\scriptsize \boldmath $\circ$}}}
\put(39,36){\makebox(0,0)[cc]{\textcolor{protozoa}{\scriptsize \boldmath $\circ$}}}
\put(116,50){\makebox(0,0)[cc]{\textcolor{protozoa}{\scriptsize \boldmath $\circ$}}}
\put(166,89){\makebox(0,0)[cc]{\textcolor{protozoa}{\scriptsize \boldmath $\circ$}}}
\put(89,117){\makebox(0,0)[cc]{\textcolor{protozoa}{\scriptsize \boldmath $\circ$}}}
\put(36,36){\makebox(0,0)[cc]{\textcolor{protozoa}{\scriptsize \boldmath $\circ$}}}
\put(174,112){\makebox(0,0)[cc]{\textcolor{protozoa}{\scriptsize \boldmath $\circ$}}}
\put(84,43){\makebox(0,0)[cc]{\textcolor{protozoa}{\scriptsize \boldmath $\circ$}}}
\put(16,59){\makebox(0,0)[cc]{\textcolor{protozoa}{\scriptsize \boldmath $\circ$}}}
\put(96,106){\makebox(0,0)[cc]{\textcolor{protozoa}{\scriptsize \boldmath $\circ$}}}
\put(97,40){\makebox(0,0)[cc]{\textcolor{protozoa}{\scriptsize \boldmath $\circ$}}}
\put(134,53){\makebox(0,0)[cc]{\textcolor{protozoa}{\scriptsize \boldmath $\circ$}}}
\put(63,78){\makebox(0,0)[cc]{\textcolor{protozoa}{\scriptsize \boldmath $\circ$}}}
\put(171,104){\makebox(0,0)[cc]{\textcolor{protozoa}{\scriptsize \boldmath $\circ$}}}
\put(24,30){\makebox(0,0)[cc]{\textcolor{protozoa}{\scriptsize \boldmath $\circ$}}}
\put(85,89){\makebox(0,0)[cc]{\textcolor{protozoa}{\scriptsize \boldmath $\circ$}}}
\put(183,33){\makebox(0,0)[cc]{\textcolor{protozoa}{\scriptsize \boldmath $\circ$}}}
\put(31,116){\makebox(0,0)[cc]{\textcolor{protozoa}{\scriptsize \boldmath $\circ$}}}
\put(95,55){\makebox(0,0)[cc]{\textcolor{protozoa}{\scriptsize \boldmath $\circ$}}}
\put(74,13){\makebox(0,0)[cc]{\textcolor{protozoa}{\scriptsize \boldmath $\circ$}}}
\put(146,8){\makebox(0,0)[cc]{\textcolor{protozoa}{\scriptsize \boldmath $\circ$}}}
\put(184,112){\makebox(0,0)[cc]{\textcolor{protozoa}{\scriptsize \boldmath $\circ$}}}
\put(77,85){\makebox(0,0)[cc]{\textcolor{protozoa}{\scriptsize \boldmath $\circ$}}}
\put(56,19){\makebox(0,0)[cc]{\textcolor{protozoa}{\scriptsize \boldmath $\circ$}}}
\put(90,87){\makebox(0,0)[cc]{\textcolor{protozoa}{\scriptsize \boldmath $\circ$}}}
\put(30,66){\makebox(0,0)[cc]{\textcolor{protozoa}{\scriptsize \boldmath $\circ$}}}
\put(110,94){\makebox(0,0)[cc]{\textcolor{protozoa}{\scriptsize \boldmath $\circ$}}}
\put(169,81){\makebox(0,0)[cc]{\textcolor{protozoa}{\scriptsize \boldmath $\circ$}}}
\put(154,19){\makebox(0,0)[cc]{\textcolor{protozoa}{\scriptsize \boldmath $\circ$}}}
\put(76,61){\makebox(0,0)[cc]{\textcolor{protozoa}{\scriptsize \boldmath $\circ$}}}

\put(197,0){\color{medium1}{\rule{290\unitlength}{   120\unitlength}}}

\put(194,80){
\thicklines
\mpt(0,0)(32,0){7}{
\put(0,0){\bezier{125}(18,-20)(40,40)(57,-20)}
	      }
\mpt(0,0)(32,0){6}{
\put(0,0){\bezier{125}(57,-20)(62,-40)(57,-50)}
\put(0,0){\bezier{125}(57,-50)(52,-53)(47,-50)}
\put(0,0){\bezier{125}(47,-50)(43,-35)(50,-20)}
		  }
\put(18, -20){\line(-1,0){6}}
\put(250,-20){\line(1,0){6}}
	 	    }
\put(193.0,1.5){\color{chicken}{\rule{15.0\unitlength}{   117.3\unitlength}}}

\put(0,0){\line(0,1){ 120}}
\put(450,0){{\rule{6\unitlength}{   75\unitlength}}}
\mpt(0,0)(0, 120){2}{\line(1,0){486}}
\mpt(-2,-2)(0, 124){2}{\line(1,0){488}}
\mpt(-4,-4)(0, 128){2}{\line(1,0){490}}
\put(-2,-2){\line(0,1){ 124}}
\put(-4,-4){\line(0,1){ 128}}

\put(15,0){
\put(0,-50){
\put(-20,-56){\color{red}{\rule{490\unitlength}{   50\unitlength}}}
	   }
\put(0,50){
\put(-20,126){\color{cyan}{\rule{490\unitlength}{   50\unitlength}}}
	  }
	  }

\mpt(50,1.5)(340,0){2}{{\rule{6.0\unitlength}{   14\unitlength}}}
\mpt(50,105)(340,0){2}{{\rule{6.0\unitlength}{   14\unitlength}}}


\put(130,90){\makebox(0,0)[cc]{\it 1}}

\put(233,-30){
\put(15,0){\makebox(0,0)[cc]{\it 2}}
{\thicklines \put(0,0){\vector(-1,2){30}} }
	   }

\put(312,-30){
\put(15,0){\makebox(0,0)[cc]{\it 3}}
{\thicklines \put(0,0){\vector(-1,2){30}} }
	   }

\put(55,150){\makebox(0,0)[cc]{\it 4}}
\put(55,-30){\makebox(0,0)[cc]{\it 4}}
\put(385,-30){\makebox(0,0)[cc]{\it 4}}
\put(385,150){\makebox(0,0)[cc]{\it 4}}
\put(240,201){\makebox(0,0)[cc]{\it 6}}
\put(240,-80){\makebox(0,0)[cc]{\it 5}}

\put(500,100){
\put(15,0){\makebox(0,0)[cc]{\it 7}}
{\thicklines \put(0,0){\vector(-1,0){30}} }
	   }

} 

}
\end{picture}
}\end{center}
\end{figure}

The Carnot heat engine is put into operation by introducing an amount of potential energy into the engine.
We will call this energy the start-up energy. For example, when the gas has the minimum volume in contact with the heater,
the spring is extended to maximum, Fig.~2A. In this case the start-up energy has the form of the potential energy of the extended spring.
The start-up energy can take other forms. For example, it can be added as a surplus pressure of the working body compared
to the environment or as the kinetic energy of the working body and the piston.

The cycle starts when the spring is extended to the utmost, while the gas occupies the smallest volume, point $a$ in Fig.~2A. The cylinder, which ideally
has an infinite heat conductivity, is put in contact with the heater at $T = T_s$. The spring starts compressing and moves the piston to the right such
as the volume occupied by the working body increases at constant temperature (the warmer isotherm of the Carnot cycle).
The working body receives heat and, together with the spring, performs mechanical work on moving the piston.
After the amount of received heat reaches $Q_s$, point $b$ in Fig.~2B, the contact with the heater is mechanically disrupted.
The gas further expands adiabatically until its temperature diminishes from $T_s$ to $T_0$, point $c$ in Fig.~2C.
At this point the working body is brought in contact with the cooler at $T = T_0$. The piston velocity in point $c$ becomes zero, the spring
starts extending and compresses the working body, which allows the latter to dispose heat to the cooler. At point $d$
in Fig.~2D the amount of disposed heat reaches $Q_0$, the cooler is detached from the cylinder and the gas continues to be compressed
adiabatically. Its temperature rises back to $T_s$, point $a$ in Fig.~2A. At this point the Carnot cycle is completed. Work $A$ performed
by the gas has taken the form of the kinetic energy of the piston and can be used outside the engine.
If this work is not taken away, it will continuously accumulate within the engine increasing the piston velocity and kinetic energy
with each cycle. In the result,
the power of the engine (i.e., the number of cycles per unit time) will increase.

Two aspects need to be emphasized. First, the start-up energy is principally important for the heat engine to operate.
If there is no spring, and the gas in the cylinder is in thermodynamic equilibrium with the heater and with the environment,
the piston will remain immobile, the engine will not operate and no work will be produced.
In the case of an infinite environment, Fig.~1A, which pressure $p$ does
not change during the cycle, the necessary amount of the start-up energy $E$ (J~mol$^{-1}$) can be calculated
as the difference between work $A_p$ performed by the piston on the environment with constant pressure $p$
and work $A_w$ performed by the working gas with $p(v) \le p$ on the piston as the piston moves from point $a$ to point $c$,
$E = A_p - A_w = \int_a^c (p - p(v))dv = p\Delta v - \int_a^c p(v)dv \ge 0$, where $v$ is molar volume and $pv = RT$ is the equation of state
for the ideal gas, $R$ is the universal gas constant. It is easy to see that
at $\Delta v/v_a \sim 1$, $\Delta v \equiv v_b - v_a$, the start-up energy $E$ should be of the order of
$Q_s = \int_a^b p(v)dv = RT_s \ln (1 + \Delta v/v_a)$.
Indeed, we have $E \ge p\Delta v - RT_s \ln(1 + \Delta v/v_a) \approx p\Delta v - RT_s(\Delta v/v_a) +
RT_s(\Delta v/v_a)^2/2 \sim RT_s/2$ at $\Delta v/v_a \sim 1$. This is a conservative estimate that ignores
work performed on the adiabate $b$-$c$ where the working body continues to expand.

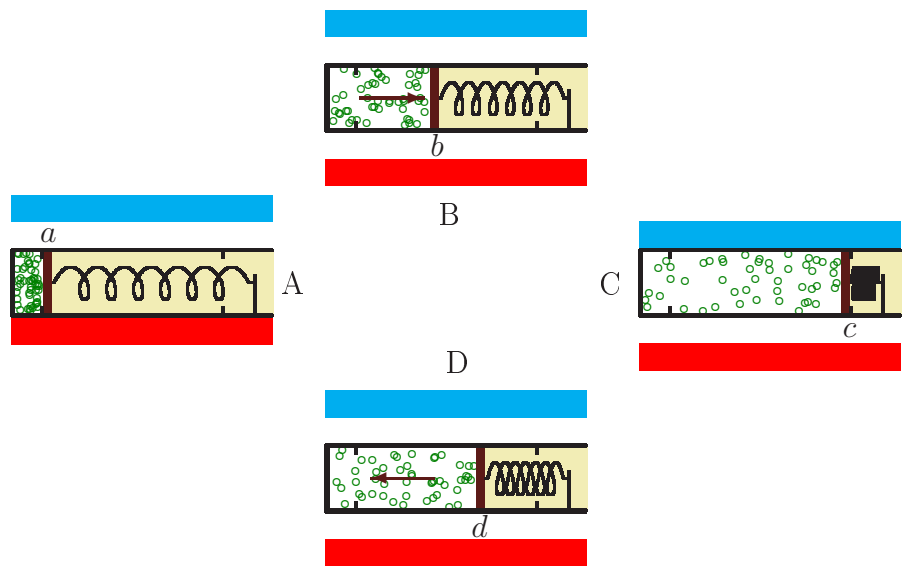
\begin{figure}
\caption{Carnot cycle. {\bf A:} beginning of the warmer isotherm at $T= T_s$, the working body is brought in contact with the heater, the spring
pulls the piston to the right, the gas expands. The initial pressure $p$ of gas at point $a$ coincides with that of the external environment.
{\bf B:} end of the warmer isotherm, beginning of the first adiabate, the heater is detached from the engine,
the gas expands adiabatically, its temperature drops from $T = T_s$ to $T = T_0$. {\bf C:} end of the first adiabate, beginning of the colder isotherm
at $T = T_0$, the working body is brought in contact with the cooler, the spring extends and pushes the piston to the left.
{\bf D:} end of the colder
isotherm, beginning of the second adiabate, gas is compressed by the moving piston, gas temperature increases from $T = T_0$ to $T = T_s$. Note
that in the end of the cycle the piston has acquired kinetic energy equal to net work $A$ performed by the working body in the cycle. To keep the engine
stationary, this energy should be taken away from the engine at point $a$.}

\newcommand{\mpt}{\multiput}

\unitlength=0.07mm
\special{em:linewidth 0.7pt}
\linethickness{0.7pt}

\begin{center}{
\begin{picture}(600.00,1200.00)(0,0)

\put(0,250){

\put(-540,250){
\put(525,60){\makebox(0,0)[cc]{\bf A}}
\put(55,150){\makebox(0,0)[cc]{ $a$}}

\put(33,16){\makebox(0,0)[cc]{\textcolor{protozoa}{\tiny \boldmath $\circ$}}}
\put(46,104){\makebox(0,0)[cc]{\textcolor{protozoa}{\tiny \boldmath $\circ$}}}
\put(41,80){\makebox(0,0)[cc]{\textcolor{protozoa}{\tiny \boldmath $\circ$}}}
\put(45,23){\makebox(0,0)[cc]{\textcolor{protozoa}{\tiny \boldmath $\circ$}}}
\put(22,90){\makebox(0,0)[cc]{\textcolor{protozoa}{\tiny \boldmath $\circ$}}}
\put(42,46){\makebox(0,0)[cc]{\textcolor{protozoa}{\tiny \boldmath $\circ$}}}
\put(46,15){\makebox(0,0)[cc]{\textcolor{protozoa}{\tiny \boldmath $\circ$}}}
\put(10,115){\makebox(0,0)[cc]{\textcolor{protozoa}{\tiny \boldmath $\circ$}}}
\put(28,101){\makebox(0,0)[cc]{\textcolor{protozoa}{\tiny \boldmath $\circ$}}}
\put(31,114){\makebox(0,0)[cc]{\textcolor{protozoa}{\tiny \boldmath $\circ$}}}
\put(41,95){\makebox(0,0)[cc]{\textcolor{protozoa}{\tiny \boldmath $\circ$}}}
\put(30,95){\makebox(0,0)[cc]{\textcolor{protozoa}{\tiny \boldmath $\circ$}}}
\put(42,20){\makebox(0,0)[cc]{\textcolor{protozoa}{\tiny \boldmath $\circ$}}}
\put(45,55){\makebox(0,0)[cc]{\textcolor{protozoa}{\tiny \boldmath $\circ$}}}
\put(14,70){\makebox(0,0)[cc]{\textcolor{protozoa}{\tiny \boldmath $\circ$}}}
\put(48,46){\makebox(0,0)[cc]{\textcolor{protozoa}{\tiny \boldmath $\circ$}}}
\put(6,43){\makebox(0,0)[cc]{\textcolor{protozoa}{\tiny \boldmath $\circ$}}}
\put(30,64){\makebox(0,0)[cc]{\textcolor{protozoa}{\tiny \boldmath $\circ$}}}
\put(44,22){\makebox(0,0)[cc]{\textcolor{protozoa}{\tiny \boldmath $\circ$}}}
\put(8,52){\makebox(0,0)[cc]{\textcolor{protozoa}{\tiny \boldmath $\circ$}}}
\put(44,31){\makebox(0,0)[cc]{\textcolor{protozoa}{\tiny \boldmath $\circ$}}}
\put(19,115){\makebox(0,0)[cc]{\textcolor{protozoa}{\tiny \boldmath $\circ$}}}
\put(31,37){\makebox(0,0)[cc]{\textcolor{protozoa}{\tiny \boldmath $\circ$}}}
\put(34,85){\makebox(0,0)[cc]{\textcolor{protozoa}{\tiny \boldmath $\circ$}}}
\put(10,66){\makebox(0,0)[cc]{\textcolor{protozoa}{\tiny \boldmath $\circ$}}}
\put(46,67){\makebox(0,0)[cc]{\textcolor{protozoa}{\tiny \boldmath $\circ$}}}
\put(7,86){\makebox(0,0)[cc]{\textcolor{protozoa}{\tiny \boldmath $\circ$}}}
\put(24,117){\makebox(0,0)[cc]{\textcolor{protozoa}{\tiny \boldmath $\circ$}}}
\put(28,10){\makebox(0,0)[cc]{\textcolor{protozoa}{\tiny \boldmath $\circ$}}}
\put(8,117){\makebox(0,0)[cc]{\textcolor{protozoa}{\tiny \boldmath $\circ$}}}
\put(7,29){\makebox(0,0)[cc]{\textcolor{protozoa}{\tiny \boldmath $\circ$}}}
\put(13,8){\makebox(0,0)[cc]{\textcolor{protozoa}{\tiny \boldmath $\circ$}}}
\put(41,57){\makebox(0,0)[cc]{\textcolor{protozoa}{\tiny \boldmath $\circ$}}}
\put(31,11){\makebox(0,0)[cc]{\textcolor{protozoa}{\tiny \boldmath $\circ$}}}
\put(20,63){\makebox(0,0)[cc]{\textcolor{protozoa}{\tiny \boldmath $\circ$}}}
\put(34,66){\makebox(0,0)[cc]{\textcolor{protozoa}{\tiny \boldmath $\circ$}}}
\put(38,69){\makebox(0,0)[cc]{\textcolor{protozoa}{\tiny \boldmath $\circ$}}}
\put(13,58){\makebox(0,0)[cc]{\textcolor{protozoa}{\tiny \boldmath $\circ$}}}
\put(33,46){\makebox(0,0)[cc]{\textcolor{protozoa}{\tiny \boldmath $\circ$}}}
\put(30,40){\makebox(0,0)[cc]{\textcolor{protozoa}{\tiny \boldmath $\circ$}}}
\put(8,17){\makebox(0,0)[cc]{\textcolor{protozoa}{\tiny \boldmath $\circ$}}}
\put(32,7){\makebox(0,0)[cc]{\textcolor{protozoa}{\tiny \boldmath $\circ$}}}
\put(36,63){\makebox(0,0)[cc]{\textcolor{protozoa}{\tiny \boldmath $\circ$}}}
\put(36,12){\makebox(0,0)[cc]{\textcolor{protozoa}{\tiny \boldmath $\circ$}}}
\put(30,97){\makebox(0,0)[cc]{\textcolor{protozoa}{\tiny \boldmath $\circ$}}}
\put(33,53){\makebox(0,0)[cc]{\textcolor{protozoa}{\tiny \boldmath $\circ$}}}
\put(18,87){\makebox(0,0)[cc]{\textcolor{protozoa}{\tiny \boldmath $\circ$}}}
\put(40,68){\makebox(0,0)[cc]{\textcolor{protozoa}{\tiny \boldmath $\circ$}}}
\put(48,57){\makebox(0,0)[cc]{\textcolor{protozoa}{\tiny \boldmath $\circ$}}}
\put(49,55){\makebox(0,0)[cc]{\textcolor{protozoa}{\tiny \boldmath $\circ$}}}

\put(56,0){\color{medium1}{\rule{431\unitlength}{   120\unitlength}}}

\put(60,80){
\thicklines
\mpt(6,0)(50.5,0){7}{
\put(0,0){\bezier{125}(9,-20)(40,40)(72.5,-20)}
	      }
\mpt(21,0)(50.5,0){6}{
\put(0,0){\bezier{125}(57,-20)(62,-40)(57,-50)}
\put(0,0){\bezier{125}(57,-50)(52,-53)(47,-50)}
\put(0,0){\bezier{125}(47,-50)(40,-35)(44.5,-20)}
		  }
\put(17, -20){\line(-1,0){6}}
\put(383,-20){\line(1,0){6}}
	    }

\put(55.5,1.5){\color{chicken}{\rule{15.0\unitlength}{   117.3\unitlength}}}

\put(15,0){
\put(-20,-56){\color{red}{\rule{490\unitlength}{   50\unitlength}}}
\put(0,50){
\put(-20,126){\color{cyan}{\rule{490\unitlength}{   50\unitlength}}}
	  }
	  }

\put(0,0){\line(0,1){ 120}}
\put(450,0){{\rule{6\unitlength}{   75\unitlength}}}
\mpt(0,0)(0, 120){2}{\line(1,0){486}}
\mpt(-2,-2)(0, 124){2}{\line(1,0){488}}
\mpt(-4,-4)(0, 128){2}{\line(1,0){490}}
\put(-2,-2){\line(0,1){ 124}}
\put(-4,-4){\line(0,1){ 128}}

\mpt(50,1.5)(340,0){2}{{\rule{6.0\unitlength}{   14\unitlength}}}
\mpt(50,105)(340,0){2}{{\rule{6.0\unitlength}{   14\unitlength}}}

} 

\put(50,600){
\put(230,-160){\makebox(0,0)[cc]{\bf B}}
\put(195,-30){\makebox(0,0)[cc]{ $b$}}

\put(88,43){\makebox(0,0)[cc]{\textcolor{protozoa}{\tiny \boldmath $\circ$}}}
\put(14,36){\makebox(0,0)[cc]{\textcolor{protozoa}{\tiny \boldmath $\circ$}}}
\put(171,55){\makebox(0,0)[cc]{\textcolor{protozoa}{\tiny \boldmath $\circ$}}}
\put(95,108){\makebox(0,0)[cc]{\textcolor{protozoa}{\tiny \boldmath $\circ$}}}
\put(174,61){\makebox(0,0)[cc]{\textcolor{protozoa}{\tiny \boldmath $\circ$}}}
\put(117,55){\makebox(0,0)[cc]{\textcolor{protozoa}{\tiny \boldmath $\circ$}}}
\put(22,31){\makebox(0,0)[cc]{\textcolor{protozoa}{\tiny \boldmath $\circ$}}}
\put(142,39){\makebox(0,0)[cc]{\textcolor{protozoa}{\tiny \boldmath $\circ$}}}
\put(102,99){\makebox(0,0)[cc]{\textcolor{protozoa}{\tiny \boldmath $\circ$}}}
\put(143,58){\makebox(0,0)[cc]{\textcolor{protozoa}{\tiny \boldmath $\circ$}}}
\put(154,12){\makebox(0,0)[cc]{\textcolor{protozoa}{\tiny \boldmath $\circ$}}}
\put(182,53){\makebox(0,0)[cc]{\textcolor{protozoa}{\tiny \boldmath $\circ$}}}
\put(11,17){\makebox(0,0)[cc]{\textcolor{protozoa}{\tiny \boldmath $\circ$}}}
\put(46,19){\makebox(0,0)[cc]{\textcolor{protozoa}{\tiny \boldmath $\circ$}}}
\put(153,84){\makebox(0,0)[cc]{\textcolor{protozoa}{\tiny \boldmath $\circ$}}}
\put(168,10){\makebox(0,0)[cc]{\textcolor{protozoa}{\tiny \boldmath $\circ$}}}
\put(41,12){\makebox(0,0)[cc]{\textcolor{protozoa}{\tiny \boldmath $\circ$}}}
\put(150,21){\makebox(0,0)[cc]{\textcolor{protozoa}{\tiny \boldmath $\circ$}}}
\put(155,106){\makebox(0,0)[cc]{\textcolor{protozoa}{\tiny \boldmath $\circ$}}}
\put(55,106){\makebox(0,0)[cc]{\textcolor{protozoa}{\tiny \boldmath $\circ$}}}
\put(144,54){\makebox(0,0)[cc]{\textcolor{protozoa}{\tiny \boldmath $\circ$}}}
\put(39,36){\makebox(0,0)[cc]{\textcolor{protozoa}{\tiny \boldmath $\circ$}}}
\put(116,50){\makebox(0,0)[cc]{\textcolor{protozoa}{\tiny \boldmath $\circ$}}}
\put(166,89){\makebox(0,0)[cc]{\textcolor{protozoa}{\tiny \boldmath $\circ$}}}
\put(89,117){\makebox(0,0)[cc]{\textcolor{protozoa}{\tiny \boldmath $\circ$}}}
\put(36,36){\makebox(0,0)[cc]{\textcolor{protozoa}{\tiny \boldmath $\circ$}}}
\put(174,112){\makebox(0,0)[cc]{\textcolor{protozoa}{\tiny \boldmath $\circ$}}}
\put(84,43){\makebox(0,0)[cc]{\textcolor{protozoa}{\tiny \boldmath $\circ$}}}
\put(16,59){\makebox(0,0)[cc]{\textcolor{protozoa}{\tiny \boldmath $\circ$}}}
\put(96,106){\makebox(0,0)[cc]{\textcolor{protozoa}{\tiny \boldmath $\circ$}}}
\put(97,40){\makebox(0,0)[cc]{\textcolor{protozoa}{\tiny \boldmath $\circ$}}}
\put(134,53){\makebox(0,0)[cc]{\textcolor{protozoa}{\tiny \boldmath $\circ$}}}
\put(63,78){\makebox(0,0)[cc]{\textcolor{protozoa}{\tiny \boldmath $\circ$}}}
\put(171,104){\makebox(0,0)[cc]{\textcolor{protozoa}{\tiny \boldmath $\circ$}}}
\put(24,30){\makebox(0,0)[cc]{\textcolor{protozoa}{\tiny \boldmath $\circ$}}}
\put(85,89){\makebox(0,0)[cc]{\textcolor{protozoa}{\tiny \boldmath $\circ$}}}
\put(183,33){\makebox(0,0)[cc]{\textcolor{protozoa}{\tiny \boldmath $\circ$}}}
\put(31,116){\makebox(0,0)[cc]{\textcolor{protozoa}{\tiny \boldmath $\circ$}}}
\put(95,55){\makebox(0,0)[cc]{\textcolor{protozoa}{\tiny \boldmath $\circ$}}}
\put(74,13){\makebox(0,0)[cc]{\textcolor{protozoa}{\tiny \boldmath $\circ$}}}
\put(146,8){\makebox(0,0)[cc]{\textcolor{protozoa}{\tiny \boldmath $\circ$}}}
\put(184,112){\makebox(0,0)[cc]{\textcolor{protozoa}{\tiny \boldmath $\circ$}}}
\put(77,85){\makebox(0,0)[cc]{\textcolor{protozoa}{\tiny \boldmath $\circ$}}}
\put(56,19){\makebox(0,0)[cc]{\textcolor{protozoa}{\tiny \boldmath $\circ$}}}
\put(90,87){\makebox(0,0)[cc]{\textcolor{protozoa}{\tiny \boldmath $\circ$}}}
\put(30,66){\makebox(0,0)[cc]{\textcolor{protozoa}{\tiny \boldmath $\circ$}}}
\put(110,94){\makebox(0,0)[cc]{\textcolor{protozoa}{\tiny \boldmath $\circ$}}}
\put(169,81){\makebox(0,0)[cc]{\textcolor{protozoa}{\tiny \boldmath $\circ$}}}
\put(154,19){\makebox(0,0)[cc]{\textcolor{protozoa}{\tiny \boldmath $\circ$}}}
\put(76,61){\makebox(0,0)[cc]{\textcolor{protozoa}{\tiny \boldmath $\circ$}}}

\put(197,0){\color{medium1}{\rule{290\unitlength}{   120\unitlength}}}

\put(194,80){
\thicklines
\mpt(0,0)(32,0){7}{
\put(0,0){\bezier{125}(18,-20)(40,40)(57,-20)}
	      }
\mpt(0,0)(32,0){6}{
\put(0,0){\bezier{125}(57,-20)(62,-40)(57,-50)}
\put(0,0){\bezier{125}(57,-50)(52,-53)(47,-50)}
\put(0,0){\bezier{125}(47,-50)(43,-35)(50,-20)}
		  }
\put(18, -20){\line(-1,0){6}}
\put(250,-20){\line(1,0){6}}
	    }

\put(193.0,1.5){\color{chicken}{\rule{15.0\unitlength}{   117.3\unitlength}}}

\put(0,0){\line(0,1){ 120}}
\put(450,0){{\rule{6\unitlength}{   75\unitlength}}}
\mpt(0,0)(0, 120){2}{\line(1,0){486}}
\mpt(-2,-2)(0, 124){2}{\line(1,0){488}}
\mpt(-4,-4)(0, 128){2}{\line(1,0){490}}
\put(-2,-2){\line(0,1){ 124}}
\put(-4,-4){\line(0,1){ 128}}

{\thicklines
\put(60,60){\color{chicken}\vector(1,0){120}}
}

\put(15,0){
\put(0,-50){
\put(-20,-56){\color{red}{\rule{490\unitlength}{   50\unitlength}}}
	   }
\put(0,50){
\put(-20,126){\color{cyan}{\rule{490\unitlength}{   50\unitlength}}}
	  }
	  }

\mpt(50,1.5)(340,0){2}{{\rule{6.0\unitlength}{   14\unitlength}}}
\mpt(50,105)(340,0){2}{{\rule{6.0\unitlength}{   14\unitlength}}}

} 

\put(640,250){
\put(-60,60){\makebox(0,0)[cc]{\bf C}}
\put(380,-30){\makebox(0,0)[cc]{ $c$}}

\put(55,89){\makebox(0,0)[cc]{\textcolor{protozoa}{\tiny \boldmath $\circ$}}}
\put(214,67){\makebox(0,0)[cc]{\textcolor{protozoa}{\tiny \boldmath $\circ$}}}
\put(362,94){\makebox(0,0)[cc]{\textcolor{protozoa}{\tiny \boldmath $\circ$}}}
\put(154,99){\makebox(0,0)[cc]{\textcolor{protozoa}{\tiny \boldmath $\circ$}}}
\put(14,40){\makebox(0,0)[cc]{\textcolor{protozoa}{\tiny \boldmath $\circ$}}}
\put(310,93){\makebox(0,0)[cc]{\textcolor{protozoa}{\tiny \boldmath $\circ$}}}
\put(153,66){\makebox(0,0)[cc]{\textcolor{protozoa}{\tiny \boldmath $\circ$}}}
\put(255,44){\makebox(0,0)[cc]{\textcolor{protozoa}{\tiny \boldmath $\circ$}}}
\put(259,26){\makebox(0,0)[cc]{\textcolor{protozoa}{\tiny \boldmath $\circ$}}}
\put(116,6){\makebox(0,0)[cc]{\textcolor{protozoa}{\tiny \boldmath $\circ$}}}
\put(367,59){\makebox(0,0)[cc]{\textcolor{protozoa}{\tiny \boldmath $\circ$}}}
\put(204,87){\makebox(0,0)[cc]{\textcolor{protozoa}{\tiny \boldmath $\circ$}}}
\put(328,19){\makebox(0,0)[cc]{\textcolor{protozoa}{\tiny \boldmath $\circ$}}}
\put(276,114){\makebox(0,0)[cc]{\textcolor{protozoa}{\tiny \boldmath $\circ$}}}
\put(175,43){\makebox(0,0)[cc]{\textcolor{protozoa}{\tiny \boldmath $\circ$}}}
\put(276,86){\makebox(0,0)[cc]{\textcolor{protozoa}{\tiny \boldmath $\circ$}}}
\put(128,101){\makebox(0,0)[cc]{\textcolor{protozoa}{\tiny \boldmath $\circ$}}}
\put(330,54){\makebox(0,0)[cc]{\textcolor{protozoa}{\tiny \boldmath $\circ$}}}
\put(222,104){\makebox(0,0)[cc]{\textcolor{protozoa}{\tiny \boldmath $\circ$}}}
\put(11,14){\makebox(0,0)[cc]{\textcolor{protozoa}{\tiny \boldmath $\circ$}}}
\put(307,21){\makebox(0,0)[cc]{\textcolor{protozoa}{\tiny \boldmath $\circ$}}}
\put(185,20){\makebox(0,0)[cc]{\textcolor{protozoa}{\tiny \boldmath $\circ$}}}
\put(47,101){\makebox(0,0)[cc]{\textcolor{protozoa}{\tiny \boldmath $\circ$}}}
\put(55,68){\makebox(0,0)[cc]{\textcolor{protozoa}{\tiny \boldmath $\circ$}}}
\put(32,88){\makebox(0,0)[cc]{\textcolor{protozoa}{\tiny \boldmath $\circ$}}}
\put(117,98){\makebox(0,0)[cc]{\textcolor{protozoa}{\tiny \boldmath $\circ$}}}
\put(296,6){\makebox(0,0)[cc]{\textcolor{protozoa}{\tiny \boldmath $\circ$}}}
\put(315,107){\makebox(0,0)[cc]{\textcolor{protozoa}{\tiny \boldmath $\circ$}}}
\put(320,48){\makebox(0,0)[cc]{\textcolor{protozoa}{\tiny \boldmath $\circ$}}}
\put(35,35){\makebox(0,0)[cc]{\textcolor{protozoa}{\tiny \boldmath $\circ$}}}
\put(139,15){\makebox(0,0)[cc]{\textcolor{protozoa}{\tiny \boldmath $\circ$}}}
\put(329,103){\makebox(0,0)[cc]{\textcolor{protozoa}{\tiny \boldmath $\circ$}}}
\put(340,99){\makebox(0,0)[cc]{\textcolor{protozoa}{\tiny \boldmath $\circ$}}}
\put(216,27){\makebox(0,0)[cc]{\textcolor{protozoa}{\tiny \boldmath $\circ$}}}
\put(220,100){\makebox(0,0)[cc]{\textcolor{protozoa}{\tiny \boldmath $\circ$}}}
\put(7,27){\makebox(0,0)[cc]{\textcolor{protozoa}{\tiny \boldmath $\circ$}}}
\put(368,100){\makebox(0,0)[cc]{\textcolor{protozoa}{\tiny \boldmath $\circ$}}}
\put(197,46){\makebox(0,0)[cc]{\textcolor{protozoa}{\tiny \boldmath $\circ$}}}
\put(80,10){\makebox(0,0)[cc]{\textcolor{protozoa}{\tiny \boldmath $\circ$}}}
\put(361,78){\makebox(0,0)[cc]{\textcolor{protozoa}{\tiny \boldmath $\circ$}}}
\put(313,23){\makebox(0,0)[cc]{\textcolor{protozoa}{\tiny \boldmath $\circ$}}}
\put(257,64){\makebox(0,0)[cc]{\textcolor{protozoa}{\tiny \boldmath $\circ$}}}
\put(364,51){\makebox(0,0)[cc]{\textcolor{protozoa}{\tiny \boldmath $\circ$}}}
\put(195,113){\makebox(0,0)[cc]{\textcolor{protozoa}{\tiny \boldmath $\circ$}}}
\put(82,64){\makebox(0,0)[cc]{\textcolor{protozoa}{\tiny \boldmath $\circ$}}}
\put(302,53){\makebox(0,0)[cc]{\textcolor{protozoa}{\tiny \boldmath $\circ$}}}
\put(241,85){\makebox(0,0)[cc]{\textcolor{protozoa}{\tiny \boldmath $\circ$}}}
\put(152,105){\makebox(0,0)[cc]{\textcolor{protozoa}{\tiny \boldmath $\circ$}}}
\put(321,83){\makebox(0,0)[cc]{\textcolor{protozoa}{\tiny \boldmath $\circ$}}}
\put(240,97){\makebox(0,0)[cc]{\textcolor{protozoa}{\tiny \boldmath $\circ$}}}

\put(375,0){\color{medium1}{\rule{112\unitlength}{   120\unitlength}}}

\put(376,80){
\mpt(-2,0)(7.5,0){7}{
\put(0,0){\bezier{125}(18,-20)(28,40)(24,-20)}
	      }
\mpt(-1,0)(7.5,0){6}{
\put(0,0){\bezier{125}(24,-20)(26,-40)(24,-50)}
\put(0,0){\bezier{125}(24,-50)(25,-53)(20,-50)}
\put(0,0){\bezier{125}(20,-50)(18,-35)(24,-20)}
		  }
\put(16, -20){\line(-1,0){6}}
\put(68,-20){\line(1,0){6}}
	    }

\put(374.5,1.5){\color{chicken}{\rule{15.0\unitlength}{   117.3\unitlength}}}

\put(15,0){
\put(0,-50){
\put(-20,-56){\color{red}{\rule{490\unitlength}{   50\unitlength}}}
	   }
\put(-20,126){\color{cyan}{\rule{490\unitlength}{   50\unitlength}}}
	    }

\put(0,0){\line(0,1){ 120}}
\put(450,0){{\rule{6\unitlength}{   75\unitlength}}}
\mpt(0,0)(0, 120){2}{\line(1,0){486}}
\mpt(-2,-2)(0, 124){2}{\line(1,0){488}}
\mpt(-4,-4)(0, 128){2}{\line(1,0){490}}
\put(-2,-2){\line(0,1){ 124}}
\put(-4,-4){\line(0,1){ 128}}

\mpt(50,1.5)(340,0){2}{{\rule{6.0\unitlength}{   14\unitlength}}}
\mpt(50,105)(340,0){2}{{\rule{6.0\unitlength}{   14\unitlength}}}

} 

\put(50,-120){
\put(245,280){\makebox(0,0)[cc]{\bf D}}
\put(275,-30){\makebox(0,0)[cc]{ $d$}}

\put(266,83){\makebox(0,0)[cc]{\textcolor{protozoa}{\tiny \boldmath $\circ$}}}
\put(105,26){\makebox(0,0)[cc]{\textcolor{protozoa}{\tiny \boldmath $\circ$}}}
\put(149,52){\makebox(0,0)[cc]{\textcolor{protozoa}{\tiny \boldmath $\circ$}}}
\put(259,57){\makebox(0,0)[cc]{\textcolor{protozoa}{\tiny \boldmath $\circ$}}}
\put(187,17){\makebox(0,0)[cc]{\textcolor{protozoa}{\tiny \boldmath $\circ$}}}
\put(52,73){\makebox(0,0)[cc]{\textcolor{protozoa}{\tiny \boldmath $\circ$}}}
\put(270,56){\makebox(0,0)[cc]{\textcolor{protozoa}{\tiny \boldmath $\circ$}}}
\put(124,15){\makebox(0,0)[cc]{\textcolor{protozoa}{\tiny \boldmath $\circ$}}}
\put(150,83){\makebox(0,0)[cc]{\textcolor{protozoa}{\tiny \boldmath $\circ$}}}
\put(159,105){\makebox(0,0)[cc]{\textcolor{protozoa}{\tiny \boldmath $\circ$}}}
\put(261,65){\makebox(0,0)[cc]{\textcolor{protozoa}{\tiny \boldmath $\circ$}}}
\put(251,57){\makebox(0,0)[cc]{\textcolor{protozoa}{\tiny \boldmath $\circ$}}}
\put(84,29){\makebox(0,0)[cc]{\textcolor{protozoa}{\tiny \boldmath $\circ$}}}
\put(84,94){\makebox(0,0)[cc]{\textcolor{protozoa}{\tiny \boldmath $\circ$}}}
\put(54,53){\makebox(0,0)[cc]{\textcolor{protozoa}{\tiny \boldmath $\circ$}}}
\put(26,42){\makebox(0,0)[cc]{\textcolor{protozoa}{\tiny \boldmath $\circ$}}}
\put(196,51){\makebox(0,0)[cc]{\textcolor{protozoa}{\tiny \boldmath $\circ$}}}
\put(33,11){\makebox(0,0)[cc]{\textcolor{protozoa}{\tiny \boldmath $\circ$}}}
\put(126,103){\makebox(0,0)[cc]{\textcolor{protozoa}{\tiny \boldmath $\circ$}}}
\put(200,54){\makebox(0,0)[cc]{\textcolor{protozoa}{\tiny \boldmath $\circ$}}}
\put(143,10){\makebox(0,0)[cc]{\textcolor{protozoa}{\tiny \boldmath $\circ$}}}
\put(211,55){\makebox(0,0)[cc]{\textcolor{protozoa}{\tiny \boldmath $\circ$}}}
\put(137,75){\makebox(0,0)[cc]{\textcolor{protozoa}{\tiny \boldmath $\circ$}}}
\put(159,65){\makebox(0,0)[cc]{\textcolor{protozoa}{\tiny \boldmath $\circ$}}}
\put(131,100){\makebox(0,0)[cc]{\textcolor{protozoa}{\tiny \boldmath $\circ$}}}
\put(275,83){\makebox(0,0)[cc]{\textcolor{protozoa}{\tiny \boldmath $\circ$}}}
\put(17,53){\makebox(0,0)[cc]{\textcolor{protozoa}{\tiny \boldmath $\circ$}}}
\put(107,48){\makebox(0,0)[cc]{\textcolor{protozoa}{\tiny \boldmath $\circ$}}}
\put(92,71){\makebox(0,0)[cc]{\textcolor{protozoa}{\tiny \boldmath $\circ$}}}
\put(158,68){\makebox(0,0)[cc]{\textcolor{protozoa}{\tiny \boldmath $\circ$}}}
\put(245,9){\makebox(0,0)[cc]{\textcolor{protozoa}{\tiny \boldmath $\circ$}}}
\put(249,85){\makebox(0,0)[cc]{\textcolor{protozoa}{\tiny \boldmath $\circ$}}}
\put(229,6){\makebox(0,0)[cc]{\textcolor{protozoa}{\tiny \boldmath $\circ$}}}
\put(64,95){\makebox(0,0)[cc]{\textcolor{protozoa}{\tiny \boldmath $\circ$}}}
\put(206,28){\makebox(0,0)[cc]{\textcolor{protozoa}{\tiny \boldmath $\circ$}}}
\put(244,37){\makebox(0,0)[cc]{\textcolor{protozoa}{\tiny \boldmath $\circ$}}}
\put(202,100){\makebox(0,0)[cc]{\textcolor{protozoa}{\tiny \boldmath $\circ$}}}
\put(199,22){\makebox(0,0)[cc]{\textcolor{protozoa}{\tiny \boldmath $\circ$}}}
\put(105,80){\makebox(0,0)[cc]{\textcolor{protozoa}{\tiny \boldmath $\circ$}}}
\put(23,101){\makebox(0,0)[cc]{\textcolor{protozoa}{\tiny \boldmath $\circ$}}}
\put(215,103){\makebox(0,0)[cc]{\textcolor{protozoa}{\tiny \boldmath $\circ$}}}
\put(220,47){\makebox(0,0)[cc]{\textcolor{protozoa}{\tiny \boldmath $\circ$}}}
\put(148,27){\makebox(0,0)[cc]{\textcolor{protozoa}{\tiny \boldmath $\circ$}}}
\put(65,24){\makebox(0,0)[cc]{\textcolor{protozoa}{\tiny \boldmath $\circ$}}}
\put(201,99){\makebox(0,0)[cc]{\textcolor{protozoa}{\tiny \boldmath $\circ$}}}
\put(60,29){\makebox(0,0)[cc]{\textcolor{protozoa}{\tiny \boldmath $\circ$}}}
\put(265,63){\makebox(0,0)[cc]{\textcolor{protozoa}{\tiny \boldmath $\circ$}}}
\put(28,114){\makebox(0,0)[cc]{\textcolor{protozoa}{\tiny \boldmath $\circ$}}}
\put(91,58){\makebox(0,0)[cc]{\textcolor{protozoa}{\tiny \boldmath $\circ$}}}
\put(197,52){\makebox(0,0)[cc]{\textcolor{protozoa}{\tiny \boldmath $\circ$}}}

{\thicklines
\put(200,60){\color{chicken}\vector(-1,0){120}}
}

\put(279,0){\color{medium1}{\rule{208\unitlength}{   120\unitlength}}}

\put(278.5,1.5){\color{chicken}{\rule{15.0\unitlength}{   117.3\unitlength}}}

\put(279,80){
\thicklines
\mpt(0,0)(19,0){7}{
\put(0,0){\bezier{125}(21,-20)(35,40)(49,-20)}
	      }
\mpt(-8,0)(19,0){6}{
\put(0,0){\bezier{125}(57,-20)(62,-40)(57,-50)}
\put(0,0){\bezier{125}(57,-50)(52,-53)(47,-50)}
\put(0,0){\bezier{125}(47,-50)(43,-35)(49,-20)}
		  }
\put(21, -20){\line(-1,0){6}}
\put(164,-20){\line(1,0){6}}
	    }
\put(15,0){
\put(0,-50){
\put(-20,-56){\color{red}{\rule{490\unitlength}{   50\unitlength}}}
	   }
\put(0,50){
\put(-20,126){\color{cyan}{\rule{490\unitlength}{   50\unitlength}}}
	  }
	   }
\put(0,0){\line(0,1){ 120}}
\put(450,0){{\rule{6\unitlength}{   75\unitlength}}}
\mpt(0,0)(0, 120){2}{\line(1,0){486}}
\mpt(-2,-2)(0, 124){2}{\line(1,0){488}}
\mpt(-4,-4)(0, 128){2}{\line(1,0){490}}
\put(-2,-2){\line(0,1){ 124}}
\put(-4,-4){\line(0,1){ 128}}

\mpt(50,1.5)(340,0){2}{{\rule{6.0\unitlength}{   14\unitlength}}}
\mpt(50,105)(340,0){2}{{\rule{6.0\unitlength}{   14\unitlength}}}

} 

}

\end{picture}
}\end{center}
\end{figure}

Second, if the heat conductivity of the heater and the cooler is sufficiently large, it ensures strict isothermy
as the piston moves from point $a$ to point $b$ at $T = T_s$ and from point $c$ to point $d$ at $T = T_0$.
Therefore, the amounts of heat $Q_s$ and $Q_0$ received and given away, respectively, by the working body are unambigously
determined by the construction of the engine. At a given $T_s$, the value of $Q_s$ is determined by the change of molar volume $v$
from point A to point B, $Q_s = \int_a^b pdv \approx p\Delta v$ for small relative changes of gas pressure $p(v)$.

The first and second laws of thermodynamics for the Carnot cycle take the form
\beq
Q_s - A = Q_0
\eeq
\beq
Q_s/T_s = Q_0/T_0
\eeq
The five magnitudes entering Eqs.~(1) and (2) leave three out of the five variables independent, e.g., $Q_s$, $T_s$ and $T_0$.
Eqs.~(1) and (2) can be written as
\beq
A = \varepsilon Q_s,\,\,\, Q_0 = (1-\varepsilon )Q_s, \,\,\, \varepsilon \equiv \frac {T_s - T_0}{T_s}.
\eeq
Here two variables are independent, $Q_s$ and $\varepsilon$ that depends on $T_s$ and $T_0$. From (3) we obtain
by replacing the independent variable $Q_s$ by $Q_0$:
\beq
A = \frac{\varepsilon}{1 - \varepsilon } Q_0,\,\,\, {\rm or}\,\,\, A = \varepsilon (Q_0 + A),\,\,\,
\frac{\varepsilon}{1 - \varepsilon} = \frac {T_s - T_0}{T_0}.
\eeq

Noteworthy, these relationships coincide with those for the dissipative heat engine if one replaces $Q_0$ by
a formally introduced variable
\beq
Q_{sd} \equiv Q_0,
\eeq
where $Q_{sd}$ refers to the external
heat input in the dissipative heat engine. Note that at $T_0 \to 0$ in Eq.~(4) work $A$ remains finite in
the view of Eq.~(2): at constant $Q_s$ and $T_s$ the decrease of $T_0$ must be accompanied
by a proportional decrease in $Q_0$.

\section{Physical meaning of the mathematical relationships of the dissipative
heat engine}

Thus, mathematics of the dissipative heat engine formally re-writes the first and second laws
of thermodynamics for the Carnot heat engine, Eqs.~(1), (2), by introducing
a new variable $Q_{sd} \equiv Q_0$. As shown in the previous section, work $A$ can be produced
by the Carnot heat engine after the engine is supplied with the start-up energy and the working
body receives heat $Q_s$ from the heater. After work is performed in the first cycle, it can be in principle dissipated
into heat at $T = T_s$ and introduced into the working body at the first stage of the second cycle.
We note that, although principally possible, such a procedure is technically difficult, as it demands
a resonance between the dynamics of the piston movement and that of the dissipation process. Characteristic
times of the piston movement and the dissipation process are dictated by different physical laws
and generally do not coincide. Resonance synchronization of these processes is in the general case impossible.

When work $A$ dissipates to heat within the working body, the latter warms. The Carnot heat engine
is originally constructed such as while the piston moves from point $a$ to point $b$, Fig.~2, the working
body receives heat $Q_s$. This is possible due to the fact that when the piston moves and
the volume of the gas increases, the gas becomes a little colder than the heater, enabling the necessary heat
to flow from the heater to the engine. If now the working body has a source of heat inside, the extension of
the working gas due to piston movement does not sufficiently decrease the gas temperature to ensure the same
flux $Q_s$ from the heater. As prescribed by the first law of thermodynamics and the ideal gas equation, gas that isothermally expands by a
preset amount (from point $a$ to point $b$ in Fig.~2) receives a fixed amount of heat $Q_s$.
If some part of this heat $Q_A = A$ is delivered to the working body as the product of dissipation of work $A$,
then the amount of external heat $Q_{sd}$ received by the engine from the heater will {\it decrease} compared
to the Carnot heat engine to $Q_{sd} = Q_s - A$. Since the amount of heat received by the working body
is invariant, the amount of work $A$ performed by the engine is also invariant and cannot be increased by dissipating
work performed by the engine in the previous cycles.

\section{The impossibility of dissipative heat engine}

Namely the central idea that "the fraction of mechanical energy dissipated... increases the heat input to the convective
heat engine" [1, p. 579] is the main physical inconsistency in the concept of the dissipative heat engine,
which brings the concept in conflict with the laws of thermodynamics.

Let $Q_{sd} = Q_s$ be the total amount of heat received by the dissipative heat engine in the first cycle (no work previously produced). Work $A_1$ produced during
this cycle is determined by the Carnot equation (3). In the subsequent cycles heat formed
due to the dissipation of work produced in the preceding cycle is added to the fixed amount of external heat $Q_{sd}$ received from the heater:
\beq
(A_d)_{n+1} = \varepsilon (Q_{sd} + (A_d)_n), \,\,\, A_1 = \varepsilon Q_{sd},
\eeq
where $n \ge 1$ is the number of the current cycle. We have from (6):
\beq
(A_d)_n  = \varepsilon (Q_s)_n,\,\,\, (Q_s)_n \equiv Q_{sd}(1 + \varepsilon + \varepsilon^2 + ... + \varepsilon^{n-1}),
\eeq
\beq
(Q_0)_n = (Q_s)_n - (A_d)_n = Q_{sd} (1 - \varepsilon^n).
\eeq
Here $(A_d)_n$ is work produced in the $n$-th cycle; $(Q_s)_n$ is the total heat received by the working body (external heat plus
the dissipative heat from work performed in the previous cycles), $(Q_s)_n$ grows with $n$; $(Q_0)_n$ is the amount of heat
given away to the cooler, it decreases with growing $n$ at $\varepsilon < 1$. As is easy to see, at $n \to \infty$
work $A_d = (A_d)_{\infty}$ is formally determined by the equation of the dissipative heat engine:
\beq
A_d = \varepsilon (Q_{sd} + A_d), \,\,\, Q_0 = (1 - \varepsilon)(Q_{sd} + A_d) = Q_{sd}.
\eeq
\beq
A_d = \frac{\varepsilon}{1 - \varepsilon} Q_{sd} = \frac{T_s - T_0}{T_0} Q_{sd}.
\eeq

In order to increase the power of a heat engine, it is necessary to increase the start-up
energy of the engine, i.e. to add potential or kinetic energy, not heat, to the engine.
Eq.~(10) says that work $A_d$ performed by the dissipative heat engine increases infinitely at fixed $Q_{sd}$
with $T_0 \to 0$. Thus, the dissipative heat engine represents an engine that re-circulates
heat to work and back at a potentially infinite power; it produces work $A_d$ greater than work
$A$ of the Carnot heat engine that operates with the same external heat input $Q_{sd} = Q_s$, cf. Eq.~(10) and Eq.~(3).
An infinite power of this re-circulation can be achieved by simply decreasing temperature $T_0$ of the cooler.
Obviously, such an engine cannot exist. Indeed, it violates both the first
and the second laws of thermodynamics. Since for a given heat engine $Q_s = Q_{sd} + A_d = {\rm const}$,
for each cycle $(A_d)_n = \varepsilon Q_s = A$ is also constant. No accumulation of energy beyond $A$ within the engine is possible in any cycle. Thus,
$A_d > A$ in the dissipative heat engine would violate the energy conservation law (the first law of thermodynamics). On the other hand,
since in the Carnot cycle all heat $Q_s$ received by the working body on the warmer isotherm is converted to work
with an entropy increase $S = Q_s/T_s \approx p\Delta v/T_s$, the condition $Q_{sd} + A_d > Q_s$ of the dissipative heat engine
would mean an additional dissipation of work $A_d$ to heat and its regeneration back to work $A_d$ from heat at one at the same temperature
(with zero entropy increment), which is prohibited by the second law of thermodynamics as formulated by Lord Kelvin.

\section{Conclusions}

Formula~(10) for work $A_d$ of a heat engine where the work produced is dissipated within the engine, is incorrect.
The problem consists in the fact that with increasing dissipation rate $(A_d)_n$ the external heat input in (6), (9)
does not remain constant, but decreases as $Q_{sd} = Q_s - (A_d)_n$, while total heat
input $Q_s$ to the working body remains constant. Putting $Q_{sd} = Q_s - A_d$ into Eqs.~(9)
gives $Q_{sd} = (1 - \varepsilon)Q_s = (T_0/T_s) Q_s$. Therefore, at $T_0 \to 0$
we have $Q_{sd} \to 0$ (no heat input from the heater) and work $A_d = A = \varepsilon Q_s = {\rm const}$ remains limited
and equal to that of a Carnot heat engine where no dissipation takes place. We summarize that dissipation of work within a heat engine {\it cannot increase the work
produced by the engine}. We also note that as far as $Q_s = Q_{sd} + A_d = {\rm const}$, the formally written entropy balance equation for the dissipative heat engine
$S_{d} = Q_{sd}/T_s + A_d/T_s - Q_0/T_0 = Q_s/T_s - Q_0/T_0 = 0$ is mathematically identical to the entropy balance equation
of the Carnot heat engine, where one term $Q_s/T_s$ is formally divided into two, $Q_s/T_s = Q_{sd}/T_s + A_d/T_s$.
The problem with the dissipative heat engine is physical, not mathematical.

Generally, the available considerations of atmospheric circulation on the basis of a thermodynamic cycle like Carnot cycle
[1-7] do not take into account the necessity of an auxilliary dynamic physical system that would possess energy $E$ and work to expand and contract the
working body of the atmospheric heat engine -- the air. The classical Carnot heat engine is based on equilibrium
thermodynamics, which prescribes that all the non-equilibrium processes of the cycle like heat transfer from the
heater to the working body occur at an infinitely small rate.
Therefore, the power (work performed per unit time) of the ideal Carnot heat engine is infinitely small.
A working body in thermodynamic equilibrium with an infinite heat source (ocean) cannot spontaneously start
expanding and receiving heat at a finite rate; this is thermodynamically prohibited.
Similarly, in the upper atmosphere there is no physical system that would compress the air and make it release
heat to the upper colder levels. Drop of air pressure observed in hurricanes cannot be the result of a Carnot heat engine
operating in the atmosphere. Patterns of atmospheric ciruclation cannot be explained on the basis of either a Carnot heat engine
or, even more so, the dissipative heat engine. The nature of atmospheric circulation is dynamic, not thermodynamic, and related to the release
of potential energy during condensation of water vapor as recently proposed [8-10].

{\bf References}

1. Renn\'{o}, N. O. and Ingersoll, A. P. Natural convection as a heat engine: A theory for CAPE. {\it J. Atmos. Sci.} {\bf 1996} {\it 53} 572-585.

2. Renn\'{o}, N. O. Reply: Remarks on natural convection as a heat engine. {\it J. Atmos. Sci.} {\bf 1997} {\it 54} 2780-2782.

3. Pauluis, O., Balaji, V. and Held, I. M. Frictional dissipation in a precipitating atmosphere. {\it J. Atmos. Sci.} {\bf 2000} {\it 57} 989-994.

4. Renn\'{o}, N. O.. Comments on "Frictional dissipation in a precipitating atmosphere". {\it J. Atmos. Sci.} {\bf 2001} {\it 58} 1173-1177.

5. Pauluis, O. and Held, I. M. Entropy budget of an atmosphere in radiativeconvective equilibrium. Part I: Maximum work and frictional dissipation.
{\it J. Atmos. Sci.} {\bf 2002} {\it 59} 125-139.

6. Bister, M. and Emanuel, K. A. Dissipative heating and hurricane intensity. {\it Meteorol. Atmos. Phys.} {\bf 1998} {\it 65} 233-240.

7. Emanuel, K. J. Tropical cyclones. {\it Annu. Rev. Earth Planet. Sci.} {\bf 2003} {\it 31} 75-104.

8. Makarieva, A. M. and Gorshkov, V. G. Biotic pump of atmospheric moisture as driver of the hydrological cycle on land. {\it Hydrol. Earth Syst. Sci.}
{\bf 2007} {\it 11} 1013-1033.

9. Makarieva, A. M. and Gorshkov, V. G. Condensation-induced dynamic gas fluxes in a mixture of condensable
and non-condensable gases. {\it Phys. Lett. A} {\bf 2009} {\it 373} 2801-2804.

10. Makarieva, A. M. and Gorshkov, V. G. Condensation-induced kinematics and dynamics of cyclones, hurricanes
and tornadoes. {\it Phys. Lett. A}, in press, doi:10.1016/j.physleta.2009.09.023.

\end{document}